# Dynamics of Collective Information Processing for Risk Encoding in Social Networks during Crises

Chao Fan[1], Fangsheng Wu[2], Ali Mostafavi[1]

[1]Zachry Department of Civil and Environmental Engineering, Texas A&M University, College Station, TX 77840 USA
[2]Department of Computer Science and Engineering, Texas A&M University, College Station, TX 77840 USA

Chao Fan, chfan@tamu.edu;
Fangsheng Wu, wufangsheng@tamu.edu;
Ali Mostafavi, amostafavi@civil.tamu.edu;

## Acknowledgements
This material is based in part upon the work supported by the National Science Foundation under Grant Number IIS-1759537, CMMI-1846069 (CAREER) and the Amazon Web Services (AWS) Machine Learning Award. Any opinions, findings, and conclusions or recommendations expressed in this material are those of the authors and do not necessarily reflect the views of the National Science Foundation and Amazon Web Services.

**Competing Interests:** The authors declare no competing interest.

## Author Contributions
All the authors designed research; C.F., and F.W. performed research and analyzed data; and C.F., and A.M. wrote the paper with comments from all authors.

## Abstract
Online social networks are increasingly being utilized for collective sense-making and information processing in disasters. However, the underlying mechanisms that shape the dynamics of collective intelligence in online social networks during disasters is not fully understood. To bridge this gap, we examine the mechanisms of collective information processing in human networks during five threat cases including airport power outage, hurricanes, wildfire, and blizzard, considering the temporal and spatial dimensions. Using the 13MM Twitter data generated by 5MM online users during these threats, we examined human activities, communication structures and frequency, social influence, information flow, and medium response time in social networks. The results show that the activities and structures are stable in growing networks, which lead to a stable power-law distribution of the social influence in networks. These temporally invariant patterns are not affected by people's memory and ties' strength. In addition, spatially localized communication spikes and global transmission gaps in the networks. The findings could inform about network intervention strategies to enable a healthy and efficient online environment, with potential long-term impact on risk communication and emergency response.

## 1. Introduction
Risks and threats are temporally evolving and spatially distributed (Banerjee et al., 2023; Liu & Fan, 2023; Xie et al., 2023). Rapidly and effectively processing situational information to inform decisions and actions is essential for people under imminently risky and uncertain conditions (Fan & Mostafavi, 2019a; Kellner et al., 2023). For decades, information processing in the face of threats was done in isolation by individuals, which mainly relied on traditional media such as radio and television, for receiving information about threats such as hurricanes and floods to make to take protective actions (Mutyebere et al., 2023). With the increased use of digital devices and social media, the communication of situational information is amplified via message transmission in online human networks (Kim et al., 2018). This has led to the emergence of collective information processing, a network process, which enables a large number of people to collectively retrieve, investigate and understand evolving threat situations (Friedland et al., 2019; Sosna et al., 2019).

Collective information processing allows people to access diverse information sources, unacquainted people suffering from the same risks (Fan, Shen, et al., 2020), as well as sentiments and reactions of these people in resistance to these threats(Zhang et al., 2019). In addition, theoretical studies such as the advances in transactive memory theory demonstrate that compared to individual information processing, the performance of people in a group would benefit from the storage, retrieval, and processing of information from other group members (Marshall et al., 2023). Despite the importance of collective information processing in social networks in threat situations, characterizing its fundamental mechanisms is still under-studied. In recent years, collective information processing has attracted attention (Shao et al., 2018), and studies have examined topics ranging from the factors for people engagement to the network effects for information adoption (Veronica et al., 2022). However, existing literature on collective information processing pays limited attention to threat situations such as crises.

Existing studies have shown that the number of people engaging in collective information processing is related to the influence of information communicators (e.g., hubs) (Yang et al., 2019). For this reason, there is one stream of research focusing on both analytical methods and empirical evidence to examine the role of information communicators. First, existing studies have developed and examined a lot of methods, such as degree centrality, betweenness centrality (Veremyev et al., 2017), PageRank (Ipsen & Wills, 2006), k-core, and HybridRank (Ahajjam & Badir, 2018) to identify influential communicators in human networks. Then, observational studies examined activity and structure patterns of information cascades that build the influence of information communicators. For example, Min et al.'s study indicates that the influence of the communicators is related to the ties that connect different communities in networks with strong modular structure (Min et al., 2015). Becker et al. presented theoretical predictions and experimental results showing that social influence generates learning dynamics that reliably improve the wisdom of crowds in processing information (Becker et al., 2017). Despite these recent advances (Fan, Esparza, et al., 2020) in uncovering the contribution of human influence to collective information processing, current research is limited to the patterns under stationary networks. In fact, human networks are dynamic with newly joined people and edges during the threats (Comfort et al., 2004). However, the roles of temporal factors related to human activities, structures, and social influence in collective information processing in threats have not been explored.

In addition to the influence of information communicators, the network effect (the effect of neighbors) is studied as an important factor for the endorsement and adoption of information in human networks also related to. To specify the network effects, a vast amount of existing studies have proposed multiple analytical pandemic models, such as susceptible-infected-recovered (SIR) model (Fan et al., 2022) and Nash equilibriums(Jackson, 2010), to approximate the real-world situations of information adoption. Moreover, recent studies (Fan & Mostafavi, 2019b) have realized the importance of the societal characteristics of people (e.g., social-economic contexts and interactions) and the interactions between these social attributes with human connections in networks. For example, Jiang et al. proposed an evolutionary game-theoretic framework to include the network users' decisions in modeling the dynamics of information diffusion(Wan et al., 2023). Akbarpour and Jackson considered the heterogeneous activity patterns across agents, and proposed and tested a method to maximize the information diffusion in classic contagion processes (Akbarpour & Jackson, 2018). These studies are generally limited to topics with large spatial coverage and people from different regions on digital platforms. However, threats and emergencies, such as hurricanes and earthquakes, tend to be localized events and affect people in a certain region. People living in different locations experience different impacts of the threats, which may cause different communication patterns and information adoption in collective information processing when facing threats in crises.

Hence, although there has been much progress in understanding the human influence and information adoption in human networks, the fundamental mechanisms related to both temporal and spatial factors in collective information processing in the context of crises threats remain a crucial missing knowledge element. In the absence of understanding of these mechanisms, the applications ranging from network modeling, forecasting to intervention for emergency response would be limited. Twitter provides a very helpful data set to capture the communication behaviors of social collectives, reflecting how events would trigger communications and instantiate social networks. In this study, the communication behaviors on Twitter include replying and re-tweeting through which people can share information with a broader audience. We use a case study of Twitter to uncover the mechanisms of risk encoding in social networks.

This study systematically examined the mechanisms of collective information processing using quantitative features considering both temporal evolution and spatial distribution factors. We project the communication behaviors as links among online social collectives to construct social networks and examine collective information processing in these networks by applying the proposed framework built upon the Network Reticulation Theory. We conducted the empirical analysis on Twitter data collected from five crises cases: (1) power outage in Atlanta airport; (2) blizzard in North American; (3) Kincade wildfire in California; (4) Hurricane Harvey in Houston; and (5) Hurricane Florence in North Carolina, with differing sizes of affected regions, populations, and extents of damages (see *Materials and Methods* for more details). We quantified the user activities, communication structures, influence scores, communication frequency, and medium response time in growing networks for each threat and examined the universality of these quantitative features across these five threat cases (Fig. 1). Each of the components in the framework are measured by the tweet activities in either temporal or spatial dimensions. Details and explanations are shared in the methods and results sections. We create the conceptual framework in Figure 1 based on the temporal and spatial dimensions. Hence, the development of the conceptual framework is based on a very precise human-social approach, and Twitter data including the user behaviors and information such as replies, tweets, location, timing, etc. allows us to quantify the components in the proposed framework.

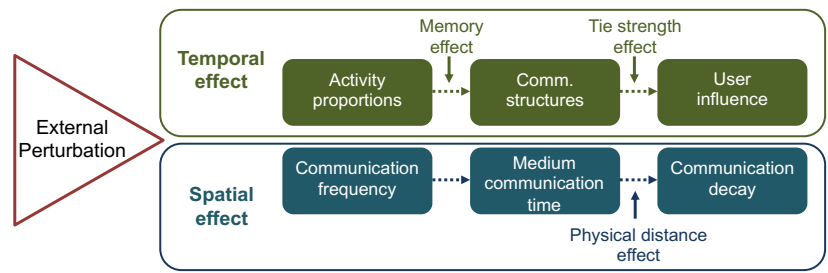

**Fig. 1.** The conceptual framework for characterizing collective information processing for encoding risks in social networks during crises.

## 2. Materials and Methods
### 2.1. Data collection and human network construction
The Twitter data used in this study were collected via the Twitter Application Programming Interface (API) and all information analyzed was publicly available. The data include the text of the messages, posted time stamps, user identifiers, location profiles, and retweet/reply/quote statuses. We selected five representative threat cases: power outage in Atlanta airport, a blizzard in North American, Kincade wildfire in California, Hurricane Harvey in Houston, and Hurricane Florence in North Carolina with differing sizes of affected regions, populations, and extents of damages. The duration of the imminent threats varies from five to ten days. To collect the data related to these selected threats, we identified relevant keywords by scanning relevant tweets when relevant warning messages for these threats were posted on Twitter. The keywords for scraping relevant tweets can be found in *Table S1 in the Supplementary Information*. Using the keywords to match the texts of the messages, we collected about 13.12 million tweets for these five threats. The number of users involved varies from 0.19 million to 2.3 million, and the relevant tweets generated during the time period vary from 0.16 million to 7.13 million. Detailed information for each threat can be found in *Table S2 in the Supplementary Information*.

Human networks are weighted directed graphs that were built upon the Twitter data for these five threats. Users who generated the tweets were extracted as the nodes. User activities including retweeting, replying, and quoting, were extracted as the directed edges, which came from the users who conducted these activities and pointed to the users who were referred in these retweets, replies or quotes. The frequencies of these activities occurring on the same pair of users were considered as the weights of the edges. These online social networks also considered the users that only posted information, but did not have any interactive activities with other users, called "isolated" nodes in graphs. To examine the mechanisms of collective information processing in growing networks, we cumulatively constructed daily human networks. That is, the edges and nodes in a daily network include all edges and nodes that appeared on this day and all the days before this day during the investigated time period. Hence, the human network on the last day of the threat period contains all edges and nodes that were generated during the data collection time period.

## 2.2. Characterization, transformation, and kernel density estimate for collective activities

Twitter allows users to conduct four types of activities, including posting original tweets, retweeting a tweet from other users and themselves, replying to a tweet to form a thread, and quoting a tweet by adding their own comments. Based on the purpose and convention of using social media, users have various tendencies of conducting different activities. The activity proportion of an individual user during the threat can indicate how the user use social media to process information and encode risks. Hence, we computed the activity proportion for each user in daily human network as follows:

$$r = \frac{Number\ of\ each\ type\ of\ activity}{all\ activties\ of\ a\ user} \quad (1)$$

Then, we examined the distribution of the users for different types of activities in different daily human networks and different threat cases. The distributions would imply the general mechanism of user activities in collective information processing. Using log-log scale, Figure 2 shows an example of user distributions for four types of activities during Hurricane Harvey. Additional results can be found in the Supplementary Information document.

To model the distributions with general statistical models, we conducted a power transformation to map the distributions into a virtual space and quantitatively characterize the distributions in that space. To do so, we first converted the activity proportion of each user using negative nature logarithmic scaling:

$$r' = -\ln r \quad (2)$$

We used negative logarithmic scaling because Box-Cox transformer requires the input data to be possible. Then, we employed the Box-Cox transformation(Box & Cox, 1964):

$$y^{(\lambda)} = \begin{cases} \dfrac{r'^{\lambda} - 1}{\lambda} & ,if\ \lambda \neq 0 \\ \ln r' & ,if\ \lambda = 0 \end{cases} \quad (3)$$

In this study, the $\lambda$ is not specified for the transformation, but the value which maximizes the log-likelihood function in this transformation will be selected. By doing so, the transformed activity proportion can be represented as a list of univariate independent and identically distributed sample observations $\left(y_1^{(\hat{\lambda})}, \dots, y_n^{(\hat{\lambda})}\right)$ from an unknown distribution, where $n$ is the number of sample observations.

The number of users involved in each threat case is large. Through the Box-Cox transformation, hence, our data can be considered as normally distributed with constant variance and expectations. To model the distribution and demonstrate the law in empirical results, we adopted a well-defined method, kernel density estimation to examine the distribution of the transformed activity proportions and show the Q-Q plot to examine the goodness-of-fit with the normal model. The kernel density estimation follows:

$$\hat{f}_h(y) = \frac{1}{nh} \sum_{i=1}^{n} K\left(\frac{y - y_i^{(\hat{\lambda})}}{h}\right) \quad (4)$$

where $\hat{f}_h(y)$ is the estimated density, $h$ is the bandwidth (in this study, we choose 0.3), $K$ is the kernel function (in this study, we choose standard Gaussian model).

## 2.3. PageRank and power-law fitting of user influence

User activities including retweet, reply, and quote are considered as a form of social endorsement or trust (Stella et al., 2018). The influence of the users in human networks can be measured by the endorsement and trust from other users in the networks. Moreover, the users become more influential if they gain endorsement from other influential users. Through this mechanism, an existing computational algorithm, PageRank, enables a quantitative measurement for user influence in human networks (Cai et al., 2023). The PageRank algorithm is based on the random walk approach which starts with a random node and move to other nodes step by step, based on the probability of an edge between them. The probability of an edge is determined by the ratio of the edge to the number of edges coming from the same source node. These edge probabilities form the initial stochastic matrix, $M_{ij}$. Due to the fragmentation of the human networks, it would be a case that all out-edges are within a group, which might cause the algorithm not to converge. To avoid this issue, at each step, a random surfer also has two options: with probability $\beta$, follows an edge at random; or with probability $1 - \beta$, jumps to a random node. The value of $\beta$ commonly ranges from 0.8 to 0.9. Then, the transition matrix is shown as follows:

$$M'_{ij} = \beta M_{ij} + (1 - \beta)\frac{1}{n} \quad (5)$$

where, $M_{ij}$ is a column stochastic matrix in which columns sum to 1; $n$ is the number of nodes (a.k.a. users) in human network. Using this transition matrix $M'_{ij}$, the PageRank of a user in human networks can be obtained as follows:

$$r = M' \cdot r \quad (6)$$

where, $r$ is the PageRank of a node. The algorithm runs until it converges (i.e., $\sum_j \left|r_j^{(t)} - r_j^{(t-1)}\right| < \varepsilon$, where $t$ is index of iteration and $\varepsilon$ is a predefined small error). This approach enables quantifying influence scores for each user who was active during the period of the threats. The distribution of user influence is plotted with 500 bins for each case. The influence scores exported from the PageRank algorithm are lower than 1, and the number of marginal users with the minimum influence scores are extremely high. This may cause expected errors in fitting the distribution. Hence, we divided the influence scores by the value of the second minimum bin, which is the lower bound of the power-law behavior in the data.

Then, we can quantify the influence distribution of human networks. In this study, we employed the maximum likelihood estimation(Clauset et al., 2009) to fit the probability density function of user influence in daily human networks with a power-law distribution ($p(x) \sim Cx^\alpha$) for the five selected threat cases. In addition, to test the goodness-of-fit, we measured the corresponding *p-values* using the Kolmogorov-Smirnov test by generating 1,000 synthetic distributions (Noulas et al., 2012). The *p-values* in this study are greater than 0.01, which indicates that the power-law model is a good fit for our data.

## 2.4. User location extraction and distance measuring

While threats tend to affect people in confined geographic region, users living in other places (e.g., other states, countries, and continents) also get involved on social media to process the risk information. To examine the interactions of users from different locations, we extract the localities of the users from user profiles. Although most of the localities are formally displayed, there are some noises induced by users' arbitrary inputs without using Twitter's official tags for their localities. Since this study aims to analyze the general mechanisms of collective information processing, we focus on the highly involved locations. Thus, for each threat, we filtered the top 100 locations based on the number of tweets generated by the users from these locations.

Twitter user locations are in a form of showing both cities and states for the locations in the United States. This form makes it easy for us to find the geo-coordinates (i.e., longitude and latitude) from Google Map Geocoding API. The distance between two locations is the great-circle distance (i.e., the shortest distance over the earth's surface) calculated using the 'haversine' formula(GIS and Geospatial Professional Community, 2017). Using the latitudes (i.e., $\varphi_A$, and $\varphi_B$) and longitudes (i.e., $\lambda_A$, and $\lambda_B$) of two localities from Google Map API, we can translate the coordinates to a distance (i.e., $d$) as follows:

$$a = sin^2\left(\frac{\varphi_B - \varphi_A}{2}\right) + cos(\varphi_A) \times cos(\varphi_B) \times sin^2\left(\frac{\lambda_B - \lambda_A}{2}\right) \quad (7)$$

$$c = 2 \times atan2\left(\sqrt{a}, \sqrt{(1-a)}\right) \quad (8)$$

$$d = R \times c \quad (9)$$

where $R = 6371000$ is the radius of the Earth in meters. Hence, the obtained distance between two locations is in meters as well.

## 3. Results

First, we examined the activities in the growing networks. The occurrence of crisis events and their impacts trigger communication activities that grows the size of online social networks related to threat topics. Our analysis is conducted on a large corpus of risk messages on Twitter which allows people to post, retweet, reply, and quote messages (Zhuang et al., 2023). These four human activities are the basis of the formation and growth of human networks for collective information processing on Twitter. Specifically, a post is an activity that enables people to report information about the situation in threats, their sentiments and reactions in response to the negative impacts. Retweets, replies, and quotes are the activities that indicate the adoption of the information posted by other people and create connections with these people. The more retweets, replies, and quotes a post receives, the more popular a post would be, and further, the more influential a person would be in the human networks. The threats triggered communication spikes on Twitter and involved a large number of people in information processing for emergency response. Examining the patterns of human activities is the first step to capture the mechanisms of collective information processing. Hence, our analysis starts with the examination of the distribution of activity proportions for each type of activity, in the case of evolving threats. The details about the number of tweets each day and the network statistics in these five threats can be found in *Supplementary Information Table S2 and Fig. S1*. The variation regarding network size and activities is generally related to the extent of damages and the scale of the affected areas.

In the top panel of Figure 2, we show the original distribution of the activity proportions. The extreme bins at the beginning and end of the distribution account for more than 90% of the activities. Those are people who only have one type of activity. As these two extreme bins are not at the same scale as other bins, the inclusion of these two bins may diminish the distribution of the rest of the people. To better characterize people who have multiple activities, we use the Box-Cox transformation for the people by excluding these extreme bins. The observed normal distribution in the bottom panel of Figure 2 can show the distribution of people with multiple behaviors and the proportion of the specific activity. The normal distribution indicates that the people of activity proportion can be estimated, and it remains consistent across the days during the disasters. Specifically, as shown in Fig. 2B, the majority of people at risk tend to retweet information from other people. It indicates that most of the people are information consumers who prefer to communicate risk information instead of generating risk information. That is because people at risk tend to actively gather information from other people to gain situation awareness before they make their decisions such as evacuation. In addition, excluding the two extreme bins at the beginning and end of the distribution, we find that the user distribution of activity proportions follows a normal distribution after Box-Cox transformation (Fig. 2) (see the goodness-of-fit in the Q-Q plots in *Supplementary Information Fig. S13-S17*). This pattern remains stable for each type of activity when the network grows day by day during Hurricane Harvey. This finding could suggest that the addition of new people and connections in human networks does not change the distribution of activities.

The activity invariance of existing and newly added people raises important questions: does this activity invariance property generally occur in other threats; if that is the case, does the activity proportion follow the same distribution in different threats? We compared the transformed probability density functions of four activities among five selected threat cases. The result demonstrates that the macroscopic user distribution of activity proportions remains stable in the growth of networks (*Supplementary Information Fig. S9-S12*). Moreover, we overlay the distributions of the activity proportions on the last day of all threats in Fig. 3. We find that the means of the distributions are almost the same, while the variances have a small different. Given this evidence, we can infer that the activity invariance and normal distribution of activities counts in transformed space are general patterns in temporally evolving social networks as collective information processing occurs for encoding risks.

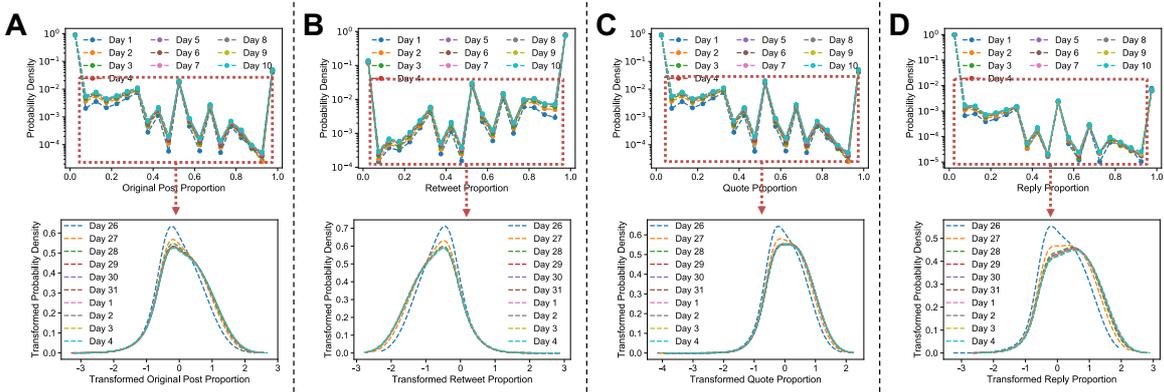

**Fig. 2.** Probability density and Box-Cox transformation of user activity proportions for different activities: original post (**A**), retweet (**B**), quote (**C**), and reply (**D**), in the case of Hurricane Harvey. (see the distributions for other threat cases in *Supplementary Materials Fig. S9-S12*.)

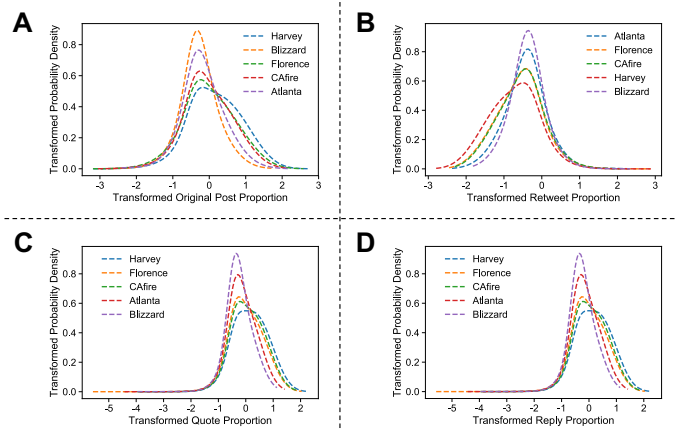

**Fig. 3.** Transformed probability density function of transformed activity proportions for different types of activities: (**A**) original post; (**B**) retweet; (**C**) quote; and (**D**) reply.

Second, we measure the structures of the networks and quantify the effect of memory on network structures. The minimum units to characterize the communication patterns and information flow are the communication structures: converging, reciprocal, and self-loop structures (Fig. 4F). The communication structures, which encode the deviation of risk information communicating from one person to another, are the direct product of human activities in the networks. This metric mainly characterizes the structure patterns of the source people who generate risk information under threats. Quantifying the patterns of communication structures is of paramount significance for both uncovering the effect of human activities for collective information processing and imposing good forecast and control of human influence in the networks (Fan, Jiang, et al., 2020).

Hence, we counted the number of users involved in each communication structures and computed the ratios in each pair of structures. As shown in Fig. 4, people, who are the information sources only involving in the converging structure, account for more than 80%, meaning the dominant role in human networks. These users only post risk information, and barely communicate or adopt information from other users. That means, the information flows from these sources to other users (i.e., information consumers), but there is very little to no situational information flow coming from the information consumers. The evidence related to the ratio of users involving in both reciprocal and converging structures also support this finding; However, the ratios are variant among different threats. In addition, the ratios are stable overall in growing networks under each threat, no matter how many new people successively participate in collective information processing. This result is consistent with the stability of the distribution of activity proportions, in which the evolution of the threats and network growth does not make effects on the distribution of activity proportions.

This finding raises another important question: is the activity and structure invariance related to the communication memory of users in threats? The communication memory of users in this context means that user's communication activities with other users is affected by their previous communications prior to the threat event. We examined the proportion of newly created edges to all edges created in each day during the threats. While the proportion of new edges decreases as the network grows, the new edges still account for more than 85% of the edges created each day (Fig. 5, and *Supplementary Information Fig. 7*). Hence, the communication memory among users does not significantly affect the activity and structure patterns during collective information processing in social networks facing threats.

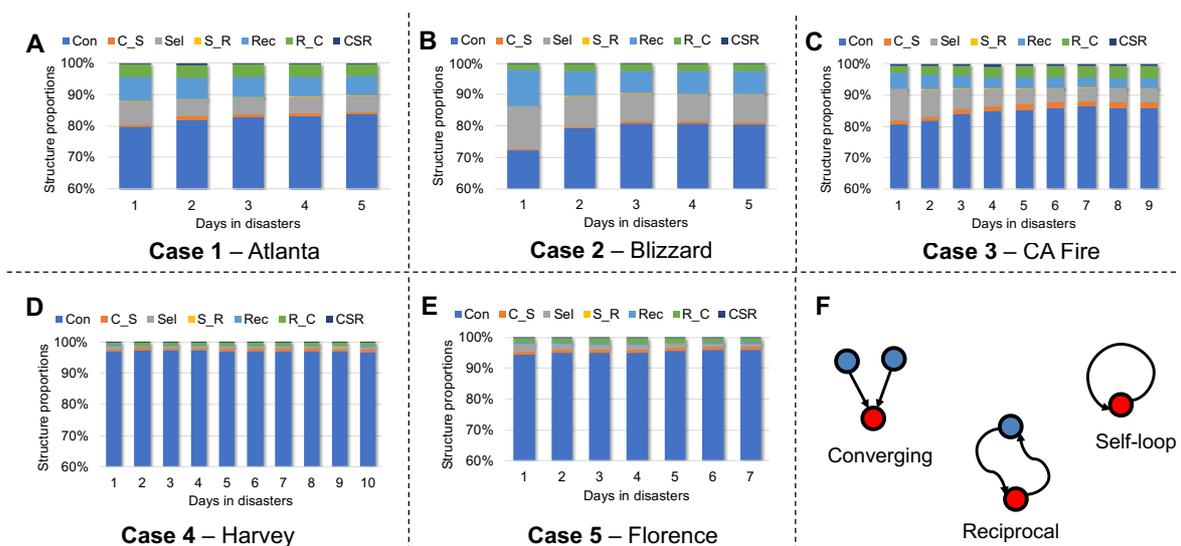

**Fig. 4.** The proportions of communication structures in growing human networks for five disaster cases: (**A**) power outage at Atlanta airport, (**B**) blizzard in North American, (**C**) Kincade wildfire in California, (**D**) Hurricane Harvey in Houston, and (**E**) Hurricane Florence in North Carolina. (**F**) Communication structures. Here, "Con" represents the users with only converging structures; "C_S" represents the users with both converging and self-loop structures; "Sel" represents the users with only self-loop structures; "S_R" represents the users with both self-loop and reciprocal structures; "Rec" represents the users with only reciprocal structures; "R_C" represents the users with both reciprocal and converging structures; and "CSR" represents the users with all three structures.

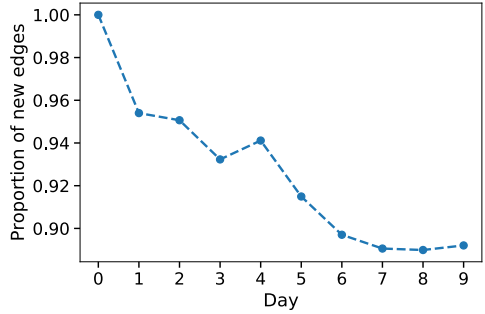

**Fig. 5.** Proportions of new edges among all edges in each day during Hurricane Harvey. (the proportions of new edges for other threats can be found in *Supplementary Information Fig. S7*)

Third, we particularly look at how the social networks are formed and what are the main characteristics in the network, including user influence, tie strength and information flows. Social networks are constructed by assembling the repeated and different communication structures. After examining the structure patterns in online social networks, we can further explore the extent to which the evolution of different threats affects the influence of users in collective information processing. Fig. 6 shows the distribution of influence scores for the people involved in communication structures (see *Materials and Methods*). The influence score is corresponding to the PageRank results for each person in the network and normalized by the medium value of the minimum bin.

Overall, we find that the distribution of the influence scores remains stable as the network grows in each threat (Fig. 6A-E). This is consistent with the pattern of structure invariance that we identified previously. The only outlier is the distribution of influence on the first day of the blizzard. The frequency of people with high influence scores is relatively lower than in other days. This might be related to the small proportion of converging structure in the network on the first day. As documented in the method of PageRank, the converging structure contributes the most to the influence of people, compared to reciprocal and self-loop structures. That is because the converging structure not only indicates the information flow from the source people to information consumers, but also implies the number of unique information consumers a source user has. Hence, the proportion of converging structure would affect the distribution of influence scores in social networks. Despite this outlier in the case of the first day, the distribution of influence reaches a stable stage starting from the second day of the blizzard. In addition, as displayed in Fig. 6F, the influence score on the last day of each threat follows the same power-law distribution. Because the power-law distribution has a heavy tail, the result implies the presence of hubs who play the role of influential communicators for generating and communicating risk information in social networks.

Yet, the PageRank algorithm does not take the weight of the edges into account. This would raise an important question: would the weights of edges affect the influence of people in networks? We investigated the weights of edges and show the distributions for each threat case (Fig. 7 for Hurricane Harvey, and *Supplementary Information Fig. S8* for the rest of the threats). We find that the medium weights of the edges for all threats are 1, and the variances are less than 3, while there are a few extreme weights that exist in the networks. Hence, in general, the edge weights do not induce variation for the influence of people in the networks. The strength of ties among the users involved in collective information processing are weak and created by people solely when they are in need of threat situation information for situation awareness.

To explore the role of influential communicators in collective information processing under threats, we selected the top 2% of people in terms of their influence scores and then examined the information flow between these influential communicators and other ordinary people in the networks. As shown in Fig. 7, we can find that while influential communicators only account for a very small percentage of the total amount of users in the network, the majority of the information propagated are originally from these influential communicators. The information generated by ordinary users and diffused by influential communicators only accounts for 1%, which is ubiquitous in all threats. This phenomenon, a signature of profound influence and information generation imbalance, is also a consequence of the power-law nature of the influence distribution(Barabási, 2009).

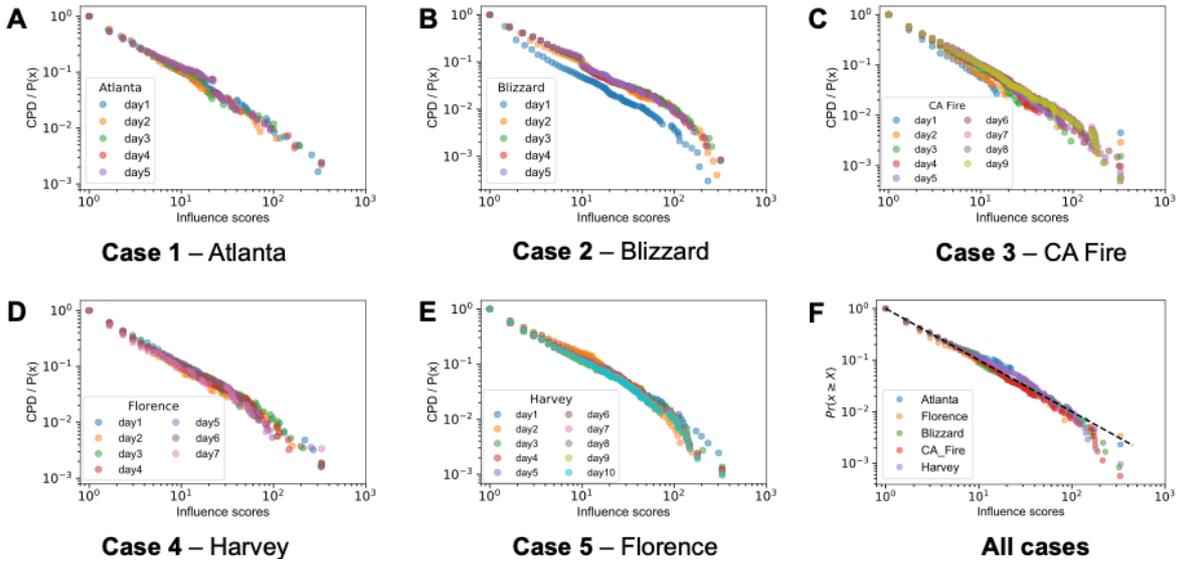

**Fig. 6.** Complementary cumulative distribution function $Pr(x \geq X)$ of normalized influence scores of users for five disaster cases: (**A**) power outage at Atlanta airport, (**B**) blizzard in North American, (**C**) Kincade wildfire in California, (**D**) Hurricane Harvey in Houston, and (**E**) Hurricane Florence in North Carolina. The influence of the marginal users who are isolated or only retweet, reply, and quote others' tweets are not included in the distribution due to the selection of $X_{min}$ for distribution fitting. (**F**) The cumulative probability density function for the whole networks in five disaster cases. The distributions follow a power law with $\alpha \approx 1.9825$.

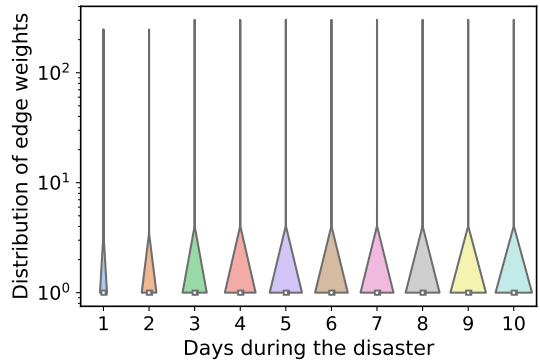

**Fig. 7.** The distribution of edge weights in each day during Hurricane Harvey. The area of each triangle represents the number of edges, and the height represents the range of possible values.

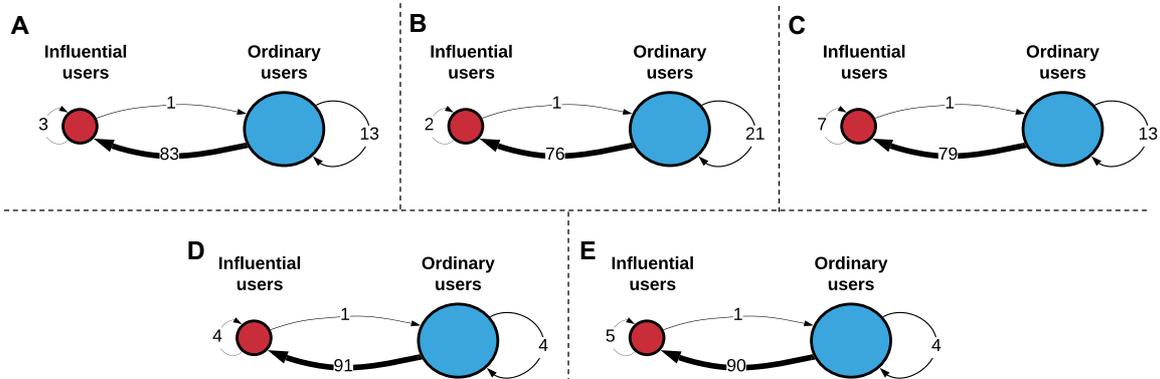

**Fig. 8.** Communication flows between influential users (top 2%) and ordinary users for five disaster cases: (**A**) power outage at Atlanta airport, (**B**) blizzard in North American, (**C**) Kincade wildfire in California, (**D**) Hurricane Harvey in Houston, and (**E**) Hurricane Florence in North Carolina. The values in the figure represent the percentages of the flow among all communication channels.

Finally, we examined the spatial patterns of the communications in social networks. Threats are spatially confined and affect people in a certain region. Although people living anywhere can participate in collective information processing on social media, the extent to which people experience and concern about the threats may vary by their proximity to threat region (Fan et al., 2021). This disparity may lead to the emergence of spatial boundaries for communications in human networks. Hence, exploring how user's locations affect the communication patterns in social networks is important for identifying and enhancing the information flow between areas. To this end, we extracted the locations of the users from the twitter messages and selected the top hundred locations in terms of the number of occurrences for our analysis. Existing studies have shown that collective human behaviors usually follow a pattern of scales (González et al., 2008). This scale pattern refers to the distance from the locations of activities to the central of neighborhoods, cities and regions (Alessandretti et al., 2020; Schläpfer et al., 2021). Based on the findings of existing studies, we use the radius as a simplified measurement for the scale of a city and a state. The location analysis uses the location information in users' profiles such as the cities, states, or geocoordinates mentioned in their profiles. The locations shared in user posts are very few, which is not suitable for our study. Although not all users have accurate location information, these data include the locations of the majority of the users.

There are a few locations (top 20 in Hurricane Harvey) where people communicate the risk information with people from other locations in a highly frequent manner (Fig. 9A, the results for other threat cases can be found in *Supplementary Information Fig. S2*). These locations are mainly in the threat-affected areas where people experience the negative impacts of the threats and tend to share the situational information with the people out of these areas. Thus, the risk information can be transmitted from the threat-affected area to these unaffected areas. However, as shown in Fig. 9A, the communication among the unaffected areas is rare. The result is more significant in the more localized threats, such as power outage in Atlanta airport and wildfire in California (*Supplementary Information Fig. S2 and S4*). The result indicates that there is information transmission decay with distance from threat-affected areas. In addition, we find that the communication spikes tend to occur among the people from the same locations (see the diagonal in Fig. 9A). This shows that people tend to adopt the information from the users in similar locations, even if the threats do not occur within their location area.

The variation of communication frequency across different locations raises an important question: is the physical distance related to the variation of communication frequency? To answer this question, we extracted the geo-coordinates for each location and computed the great-circle distance (i.e., the shortest distance over the earth's surface) for each pair of locations (see *Supplementary Information Fig. S3*). As shown in Fig. 9B, people from the same area communicate frequently. Since the locations that we identified are at a city level, there is no data corresponding to the physical distance which is shorter than the radius of a city (e.g., the distance is shorter than 2 kilometers). Beyond the scale of a city, but within the scale of a state (e.g., the distance is longer than 2 kilometers, but shorter than 100 kilometers), the communication frequency remains relatively high, meaning that people in nearby cities also concern about the situation in the threat-affected areas. However, as the distance exceeds the scale of a state and continues increasing, the communication frequency decreases gradually. This is a consequence of the nature of threats, since the impacted area is limited, and people communicate only in need of situational information under the threats. People living in the area far away from the threats might not need any information about the threats.

In addition to the communication frequency, the response time is another metric to measure the effect of location on collective information processing in threats. The response time represents the length of the period between the time when the tweet was created and the time when a person retweet/reply/quote it. Since the response time is a measure for each activity, there should be a large variation for the values of this metric. To get an objective value for this metric and reduce the effects of extreme values, we use the medium response time as the aggregated measure for this metric. As shown in Fig. 10, the pattern of the medium response time is consistent with the findings regarding the communication frequency. The results for other threat cases can be found in *Supplementary Information Fig. S5 and S6*. We find that the people living in the same residential areas (whether the area is affected by treats or not) respond to each other very fast than the response between people living in different unaffected areas (Fig. 9A). Also, people living in nearby cities respond to each other as fast as the people from the same cities.

However, the medium response time increases, as the physical distance exceeds the scale of a state and continues increasing.

To better illustrate the pattern of spatially bounded communication frequency and medium response time, we proposed a surrogate model that shows the communication frequencies (Fig. 9C) and the inverse of medium response time (Fig. 10C) between different areas using the edges with various widths. The surrogate model includes the three findings identified previously from the results regarding each communication metric.

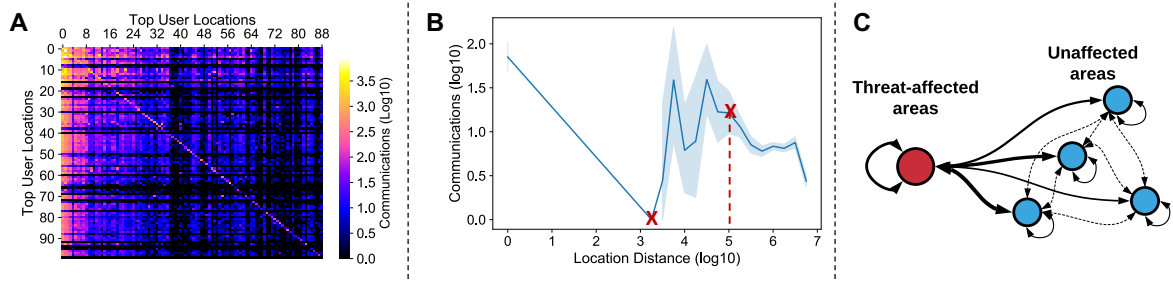

**Fig. 9.** (**A**) The communication frequency distribution across top 100 locations identified from users' profiles for Hurricane Harvey. (**B**) The relationship between communication frequency and location distances (in meters) during Hurricane Harvey. The first mark indicates the scale of a city, and the second mark indicates the scale of a state. (**C**) A surrogate model for an illustration of the spatially bounded communication frequencies in threats (the width of the edges represents the communication frequency).

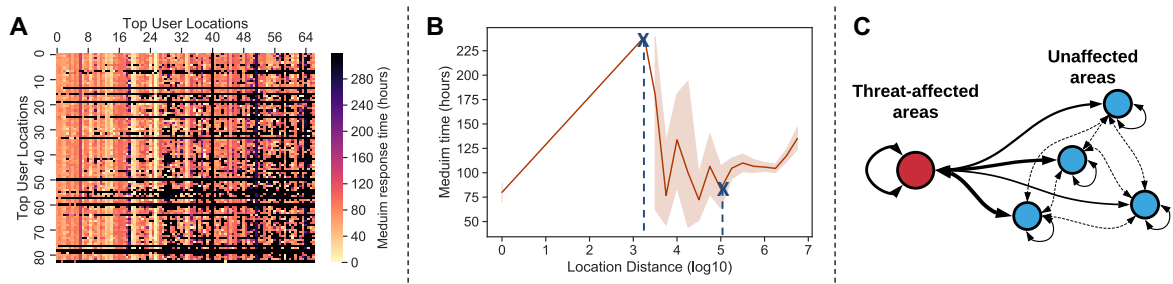

**Fig. 10.** (**A**) The distribution of medium response time across top 100 locations identified from users' profiles for Hurricane Harvey. (**B**) The relationship between medium response time and location distances (in meters) during Hurricane Harvey. The first mark indicates the scale of a city, and the second mark indicates the scale of a state. (**C**) A surrogate model for an illustration of the spatially bounded medium response time in threats (the width of the edges represents the inverse of the medium response time).

## 5. Conclusion.

This study provides quantitative empirical evidence needed to characterize the fundamental and universal mechanisms of collective information processing for risk encoding in social networks during crises. Specifically, the risk-encoded social networks are temporally invariant in terms of their network structures, user activities, and influence scores. Therefore, throughout the period of crises, influential communicators always play a primary role in generating and communicating information and bridging gaps between ordinary people. This information, due to its criticality to local events and population, is usually spatially bounded. The results shown in this work have both theoretical and practical implications for network modeling and interventions. This study focuses on characterizing the mechanisms of collective information processing under threat conditions. The present finding supports the application of existing network models considering the temporal stability and spatial boundary in rapid-evolving situations such as those studied throughout this work. This study is in response to the increasingly restrictive policies on the use of these social media platforms and the pressure on their business model. The findings could help policymakers in setting up legal obligations and fixing responsibilities for these platforms.

The findings presented in this study suggest three strategies to enable an efficient online environment for collective information processing in the face of threats. Under risky and uncertain conditions, the efficiency of gathering situational is critical for people to take protective actions and make a rapid response. Response agencies, such as government officials, volunteers, and relief organizations who have relief resources but are not influential on social media, can request for transmission by influential communicators. Second, making an effective response not only relies on the speed of information gathering, but also relates to the credibility of the received situational information. To examine the information credibility, the influence scores of the communicators might provide a useful signal to prioritize accounts for further fact-checking. Finally, if the mis-/dis-information has been communicated in human networks, curbing influential communicators who involve in generating and transmitting this information would be an effective strategy to mitigate the negative impact of this mis-/dis-information. That is because influential communicators, despite only a small proportion to the entire network, play a vital role in connecting ordinary people. The network is robust to random selection and removal of nodes, but frail to the removal of influential communicators.

Finally, the communication frequency and response time decay with the physical distances between users in the networks. People are more likely to communicate with the users in the same city. Therefore, people's location becomes a factor that groups users in small clusters, in which users are more cohesive with each other. This mechanism would lead to efficient information processing among people from the same region and impede the information exchange between two clusters (i.e., regions). The information processing could be benefit from the cross-boundary edges created by the people active in different clusters. Combining the previous finding that influential communicators play the primary role in information processing and communication, an effective strategy would be encouraging the connections between influential users from different spatial clusters. Progress in this direction may be accelerated through developing recommendation algorithms for influential users on social media platforms. For example, social media platforms can recommend popular risk messages from a cluster to an influential communicator in another cluster (Anwar et al., 2023). This strategy could contribute to requesting relief resources such as donations from the people in unaffected regions.

This work has some limitations and needs future work to improve. First, all social networks studied in this paper are in crises and in the same country. Considering the cultural differences and socio-demographic backgrounds, it would be important to explore whether the identified mechanisms are also applicable in regular circumstances, and whether the mechanisms would change in other countries and social media platforms. For example, one question that can be further addressed is whether similar patterns of collective communication might be observed in processing the information of political news and economic events. It would also be helpful to consider other types of events such as festivals or with other social networks to conduct a comparison analysis. Standard datasets from TREC, SEMEVAL, and CLEF forums could be considered. Second, the social network data employed in this study are only Twitter data, as Twitter has been provided free API access for academic research. However, social networks are far beyond just Twitter platforms. Many other platforms and places such as message communications on mobile phones, physical contact in stores, interactions on Facebook, etc. However, collecting fine-grained data sets from different platforms to construct comprehensive social networks is very challenging for research community. The limited access to different data sets may hinder the depth of understanding on collective behaviors in social networks. Third, social network data tend to be very large, accessing and processing the data still requires programming skills, which may be a limitation of broadly using these data for research. Mainstream social media such as Meta and Instagram also allow people to manipulate information with multiple functionalities. However, the societal characteristics (e.g., age, gender, and culture) of the users on different social media might be different and induce potential impacts on human activities as well as network structures. Hence, systematic analysis of the mechanisms of collective information processing on different platforms deserves further investigation. Fourth, this study focuses on understanding how the information is endorsed and adopted by other users. How the information is spread and appears in other users' feeds is not included in our analyses. The number of followers and the recommendation algorithms may influence the spread of the information, which may further influence the patterns of collective information processing. It would be interesting to see the examination of these factors in future studies. Finally, a user survey is helpful to further explain and validate the findings of our studies. Although it is very challenging to reach out to the users or conduct a user survey, cross-checking users' motivations would be helpful to deepen our understanding of collective information processing.